\documentclass[letterpaper, twocolumn,9pt]{article} 
\usepackage[T1]{fontenc}  

\usepackage{newtxtext,newtxmath} 
\usepackage{titlesec}
\titleformat{\subsection}
  [runin]                 
  {\bfseries}             
  {}                      
  {0pt}
  {}                      
  [.]                     
\titlespacing*{\subsection}
  {0pt}                   
  {6pt}                   
  {0.5em}                 

\usepackage{geometry} 
\usepackage{setspace}

\usepackage[
    backend=biber,
    style=chem-acs,
    citestyle=numeric-comp,
    sorting=none,
    doi = true,
    maxbibnames=5
]{biblatex}  
\usepackage{graphicx}
\usepackage{float}
\newfloat{scheme}{htbp}{los}
\floatname{scheme}{Scheme}
\floatname{chart}{Chart}
\newfloat{graph}{htbp}{loh}

\usepackage{chemformula} 
\usepackage[version = 4]{mhchem} 
\usepackage{upgreek} 
\usepackage{units} 
\usepackage{multirow} 
\usepackage{placeins} 
\usepackage{cuted} 

\usepackage{xcolor} 
\definecolor{blue_hyperref}{HTML}{0D54A6}

\definecolor{notegray}{HTML}{A0A0A0}

\usepackage{hyperref} 
\hypersetup{
    colorlinks=true,
    linkcolor=blue_hyperref,
    citecolor=blue_hyperref,
    filecolor=blue_hyperref,
    urlcolor=blue_hyperref,
    pdftitle={Friedel et al.},
}

\usepackage[nameinlink]{cleveref} 
\usepackage{subcaption}

\newcommand{\eqsMr}{M_\mathrm{r}} 
\newcommand{\eqsMs}{M_\mathrm{sat}} 
\newcommand{\eqsHext}{H_\mathrm{ext}} 
\newcommand{\eqsHwidth}{H_\mathrm{width}} 
\newcommand{\eqsomegaext}{\omega_\mathrm{H}} 
\newcommand{\eqsomegaexch}{\omega_\mathrm{A}} 
\newcommand{\eqsomegaM}{\omega_\mathrm{M}} 

\usepackage{xspace}
\newcommand{\muBLS}{$\upmu\mathrm{BLS}$\xspace}

\usepackage{verbatim}

\usepackage{authblk}
\author[1,2]{Anna M. Friedel*}
\author[2]{Jaafar Ghanbaja}
\author[1]{Björn Heinz}
\author[1]{Moritz Bechberger}
\author[2]{Sylvie Migot}
\author[2]{Sébastien Petit-Watelot}
\author[2]{Stéphane Andrieu}
\author[1]{Philipp Pirro**}
\affil[1]{Fachbereich Physik and Landesforschungszentrum OPTIMAS, Rheinland-Pfälzische Technische Universität Kaiserslautern-Landau, 67663 Kaiserslautern, Germany}
\affil[2]{Institut Jean Lamour, UMR CNRS 7198, Université de Lorraine, 54000 Nancy, France}
\newcommand{\mytitle}{Epitaxial \ch{Co2MnSi} with intrinsic magnetocrystalline anisotropy as a route to bias-field-free nonlinear half-metal magnonics at the nanoscale}
\title{\mytitle}
\date{Email: *friedel@rptu.de, **ppirro@rptu.de}

\begin{document}
    \twocolumn[
\begin{@twocolumnfalse}
\maketitle

\begin{abstract}
  Half-metallic Heusler compounds like \ch{Co2MnSi} allow to bridge magnonic and spintronic functionality for hybrid unconventional computing approaches with sought-after properties like 100\% spin polarization and associated low Gilbert damping $\alpha\leq 10^{-3}$. However, the desirable material parameters are inherently tied to the crystal lattice with a particularly critical dependence on structural order in \ch{Co2MnSi}. To date, the successful fabrication of nanoscale devices with robust structural integrity remains yet a challenge, and consequently the impact of the material parameters on the resulting nonlinear spin-wave dynamics remains largely unexplored. Here, we report on a study of linear and nonlinear spin-wave dynamics in transversally magnetized \ch{Co2MnSi} waveguides with impeccable crystalline ordering. We show that epitaxial, L2\textsubscript{1}-ordered \ch{Co2MnSi} exhibits an intrinsic cubic anisotropy with first- and second-order contributions, stabilizing a magnetization alignment along the crystal $\langle110\rangle$ directions. We confirm the implication of an unaffected crystal structure resulting in preserved magnetic properties in the patterned structures. Herein, the persistent magnetocrystalline anisotropy reshapes the spin-wave dispersion which yields a first-order nonlinear instability suppression range extending over several GHz - even for vanishing bias fields. Moreover, the intrinsic magnetocrystalline anisotropy can be exploited to counteract shape demagnetization for a stabilized low bias field operation in the favourable Damon-Eshbach geometry with high group velocities and decay lengths. Together with the proven half-metallicity and ultralow Gilbert damping, this research establishes \ch{Co2MnSi} as a robust, scalable platform towards bias-field-free nonlinear half-metal magnonics. 
\end{abstract}

\subsection*{Keywords}

half-metal magnonics, nanofabrication robustness, cubic anisotropy, nonlinear spin waves, Heusler, \ch{Co2MnSi}

\vspace{0.5cm}
\end{@twocolumnfalse}
]

\pdfbookmark[1]{Introduction}{Introduction}
Nonlinear magnetization dynamics are increasingly exploited in spintronics and magnonics, largely motivated by an application towards unconventional computing \supercite{torrejonNeuromorphicComputingNanoscale2017, grollierNeuromorphicSpintronics2020, finocchioPromiseSpintronicsUnconventional2021, pappNanoscaleNeuralNetwork2021}. The desirable combination of both approaches in hybrid devices with magnonic and spintronic functionality has long been limited by functionality compromises due to the restricted material compatibility, particularly with the magnonic "holy grail" YIG \supercite{pirroAdvancesCoherentMagnonics2021, barman2021MagnonicsRoadmap2021, flebus2024MagnonicsRoadmap2024}. Half-metal magnonics promises to bridge both fields by exploiting the unique properties of half-metallic ferromagnets \supercite{degrootNewClassMaterials1983, ishidaSearchHalfMetallicCompounds1995}, characterized by a full band gap at the Fermi energy for only one spin type. This implies a 100\% spin polarization as sought-after in spintronic devices and an associated ultralow Gilbert damping \supercite{liuOriginLowGilbert2009} as is key to magnonic applications. Specifically, the Heusler compound \ch{Co2MnSi} has recently regained increasing attention linked to the experimental proof of half-metallicity \supercite{jourdanDirectObservationHalfmetallicity2014, andrieuDirectEvidenceMinority2016} and the subsequently reported record-low associated Gilbert damping $\alpha\approx4\times10^{-4}$ for conductive thin films \supercite{guillemardUltralowMagneticDamping2019}. The compound offers a unique set of properties for an exploitation at the magnonic-spintronic intersection \supercite{pirroAdvancesCoherentMagnonics2021, barman2021MagnonicsRoadmap2021, flebus2024MagnonicsRoadmap2024}: The full spin polarization promises maximal efficiency for spintronic manipulation \supercite{sakurabaHugeSpinPolarizationL212005, chudoSpinPumpingEfficiency2011, liuGiantTunnelingMagnetoresistance2012, sasakiSpinInjectionEfficiency2020}, ensured over a large temperature range by the high Curie temperature $T_\mathrm{C}=985\,\mathrm{K}$ \supercite{websterMagneticChemicalOrder1971}. The combination of ultralow magnetic damping $\alpha$ and large saturation magnetization $\eqsMs\approx1000\,\mathrm{kA/m}$ \supercite{websterMagneticChemicalOrder1971, guillemardEngineeringCo2MnAlxSi1Heusler2020, demeloUnveilingTransportProperties2021} implies long spin-wave lifetimes and high group velocities enabling for large coherence and propagation lengths \supercite{stucklerSpinWavePropagation2018}, and the intrinsic cubic magnetocrystalline anisotropy even allows for reconfigurable device functionality in zero bias fields \supercite{mantionCubicAnisotropyReconfigurable2022,mantionReconfigurableSpinWave2024}. This is particularly favourable towards nonlinear dynamic applications, where large $\eqsMs$ and ultralow $\alpha$ promise reduced nonlinear pumping thresholds, and the cubic anisotropy not only promises a potential bias-field-free operation but also yields an additional toggle on the nonlinear instability threshold \supercite{pattonSpinwaveInstabilityTheory1979, pattonSpinWaveInstabilityTheory1979a, sekiguchiSpinwavePropagationCubic2017}. Moreover, a stable functionality is ensured down to cryogenic temperatures \supercite{guillemardUltralowMagneticDamping2019, demeloUnveilingTransportProperties2021}, where the persistently low magnetic dissipation yields a particularly timely interest in view of quantum magnonic applications \supercite{serhaMagneticMaterialsQuantum2026}. 

The great promise of \ch{Co2MnSi} for half-metal magnonics comes with one major caveat: Ensuring an impeccable sample quality is a crucial prerequisite for achieving the desired properties in Heusler compounds \supercite{felserHeuslerAlloysProperties2016, palmstromHeuslerCompoundsSpintronics2016, wollmannHeusler40Tunable2017}, and in the case of \ch{Co2MnSi} specifically requiring a precise stoichiometry as well as a structural integrity of the Heusler lattice with characteristic chemical L2\textsubscript{1} order \supercite{gaierInfluenceL21Ordering2008, kubotaStructureExchangeStiffness2009, wustenbergSurfaceSpinPolarization2012, hasnipEffectCobaltSublatticeDisorder2014, abdallahStructuralMagneticProperties2016, abdallahEvolutionMagneticProperties2017, shawMagneticDampingSputterdeposited2018, guillemardPolycrystallineCo2MnbasedHeusler2019, saitoMagneticThermoelectricProperties2021}. The remarkable structural criticality is reflected by the decades of massive research efforts \supercite{ritchieMagneticStructuralTransport2003, schmalhorstInterfaceStructureMagnetism2004, sakurabaHugeSpinPolarizationL212005, kallmayerMagneticPropertiesCo22006, chudoSpinPumpingEfficiency2011, grafSimpleRulesUnderstanding2011, liuGiantTunnelingMagnetoresistance2012} preceding the only recently reported experimental validation of half-metallicity \supercite{jourdanDirectObservationHalfmetallicity2014, andrieuDirectEvidenceMinority2016} and associated ultralow Gilbert damping \supercite{guillemardUltralowMagneticDamping2019} in \ch{Co2MnSi}. Overcoming this bottleneck has been enabled by advances in state-of-the-art epitaxial growth \supercite{guillemardIssuesGrowingHeusler2020}, now allowing for reproducible high-quality \ch{Co2MnSi} thin films with reliable stoichiometric control and predominant L2\textsubscript{1} order.

Nevertheless, the successful fabrication of nanoscale \ch{Co2MnSi} devices with preserved material properties remains an unsolved challenge. The required epitaxial growth imposes a device fabrication in a top-down approach, typically involving a patterning procedure with lithography and milling steps. However, such downscaling processes are known to affect the performance of magnonic devices \supercite{jungfleischSpinWavesMicrostructured2015, kiechleSpinWaveOpticsYIG2023, greilEffectGaionIrradiation2025}, due to the risk of damage on the crystal structure introduced for instance by atom displacement or ion implantation. Especially in the specific case of \ch{Co2MnSi}, the crucial dependence of the key parameters on the stoichiometry \supercite{guillemardPolycrystallineCo2MnbasedHeusler2019} and intact L2\textsubscript{1}-ordered Heusler lattice \supercite{abdallahStructuralMagneticProperties2016, abdallahEvolutionMagneticProperties2017} translates to a particular criticality for ensuring the desired magnonic functionality in patterned \ch{Co2MnSi} \supercite{mantionInfluenceGaMilling2022}. 

In consequence, experimental studies on the magnonic functionality of \ch{Co2MnSi} remain scarce. The majority of magnonic studies on Heusler compounds focus on the half-metallic candidate Co\textsubscript{2}Mn\textsubscript{0.6}Fe\textsubscript{0.4}Si (CMFS) \supercite{sebastianLowdampingSpinwavePropagation2012, pirroNonGilbertdampingMechanismFerromagnetic2014, sebastianChapter13Co2Mn06Fe04Si2016, langerParameterfreeDeterminationExchange2016, stucklerUltrabroadbandSpinwavePropagation2017, mallickTunabilityDomainStructure2019}. While these efforts corroborate the potential importance to the field of magnonics, a significant drawback in CMFS are the unconfirmed half-metallicity as well as the significant non-Gilbert damping contribution revealed by peculiar nonlinear dynamics \supercite{pirroNonGilbertdampingMechanismFerromagnetic2014}. Concerning \ch{Co2MnSi} on the other hand, experimental studies on the magnonic functionality remain limited to the spin-wave propagation over large distances in macroscopic patches \supercite{stucklerSpinWavePropagation2018}, the bias-field-free operation of a magnonic crystal \supercite{mantionReconfigurableSpinWave2024} and the current-induced change of the spin-wave attenuation in macroscopic spin-wave Doppler shift devices \supercite{cordova2026Spinpolarization}. These reports underline the promising potential towards a perspective bias-field-free nonlinear device operation in magnonic circuits, highlighting the need for a non-destructive scalability of \ch{Co2MnSi} and an evaluation of the exploitable nonlinear dynamics at vanishing bias fields. 

In this work, we address three central questions for the realization of half-metal magnonics based on \ch{Co2MnSi}: (i) Can the structural and magnetic properties of epitaxial, L2\textsubscript{1}-ordered films be preserved under nanoscale top-down fabrication? (ii) How does the intrinsic cubic magnetocrystalline anisotropy shape the spin-wave dynamics in the patterned structures? And (iii) can this anisotropy be harnessed to enable stable and controllable nonlinear dynamics at vanishing bias fields? By systematically studying nanostructured waveguides derived from a single high-quality film, we demonstrate not only the preservation of the crystal structure upon nanofabrication, but explicitly confirm its direct implication in maintaining the magnetic key parameters. We further reveal the decisive role of cubic anisotropy in shaping the spin-wave dispersion and nonlinear response, and show that it enables stabilized operation in near-zero bias fields. These results establish the essential conditions for a robustly scalable magnonic functionality in \ch{Co2MnSi}.

\begin{figure*}[ht!]
  \centering
  \includegraphics{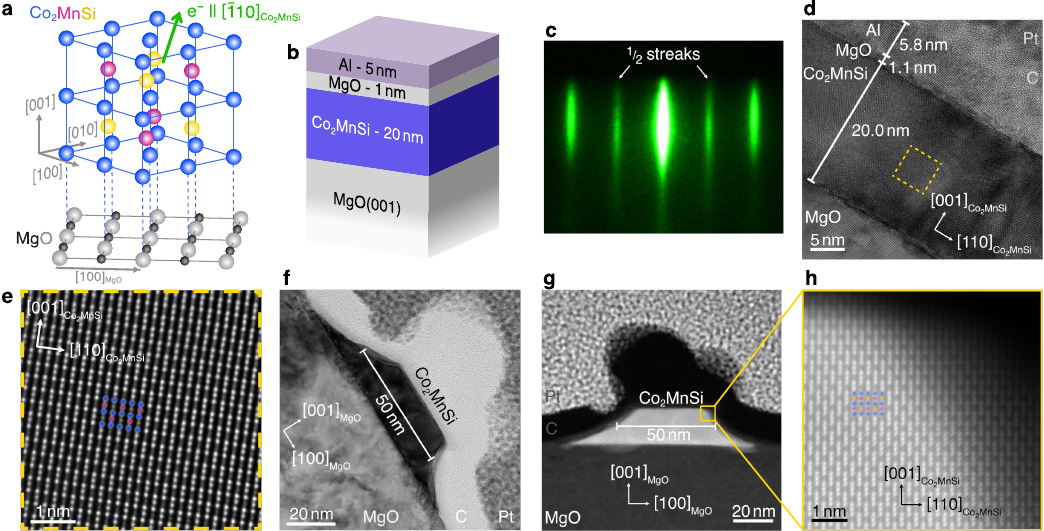}
  \caption[Nanoscale integrity of the L2\textsubscript{1}-ordered Heusler crystal structure upon nanofabrication.]{
    \textbf{Nanoscale integrity of the L2\textsubscript{1}-ordered Heusler crystal structure upon nanofabrication.} 
    \textbf{(a)} Illustration of the Heusler lattice in the L2\textsubscript{1} order on an \ch{MgO}(001) substrate \supercite{guillemardIssuesGrowingHeusler2020}, with an indication of the electron beam probing direction. \textbf{(b)} Deposited thin film stack with nominal thicknesses grown by MBE. 
    \textbf{(c)} \textit{in situ} RHEED for $\mathrm{e}^-\parallel\left[110\right]_\mathrm{\ch{Co2MnSi}}$, where the appearance of $\nicefrac{1}{2}$-streaks indicates a high chemical ordering degree. 
    \textbf{(d)} HRTEM micrograph along the $\left[\bar{1}10\right]$ zone axis of \ch{Co2MnSi} showing the epitaxial \ch{Co2MnSi}/\ch{MgO}/Al thin film stack prior to nanofabrication. The dashed yellow square indicates the size of the region imaged in \textbf{e}. 
    \textbf{(e)} Filtered HAADF-STEM micrograph revealing the characteristic intensity pattern of the L2\textsubscript{1} order as marked by the coloured atoms. 
    \textbf{(f)} HRTEM micrograph along the $\left[1\bar{1}0\right]$ zone axis of \ch{Co2MnSi} showing a cross-section of the smallest patterned waveguide with a core width of $w=50\,\mathrm{nm}$. 
    \textbf{(g)} HAADF-STEM micrograph of the same waveguide. 
    \textbf{(h)} Filtered HAADF-STEM micrograph showing a zoom-in on the top corner of the waveguide, where the characteristic intensity pattern confirms the L2\textsubscript{1}-order even in the edges of the nanostructures.
  }
  \label{fig:1_structural_integrity}

  \begin{subfigure}{0pt}
    \phantomcaption
    \label{fig:1a}
  \end{subfigure}
  \begin{subfigure}{0pt}
    \phantomcaption
    \label{fig:1b}
  \end{subfigure}
  \begin{subfigure}{0pt}
    \phantomcaption
    \label{fig:1c}
  \end{subfigure}
  \begin{subfigure}{0pt}
    \phantomcaption
    \label{fig:1d}
  \end{subfigure}
  \begin{subfigure}{0pt}
    \phantomcaption
    \label{fig:1e}
  \end{subfigure}
  \begin{subfigure}{0pt}
    \phantomcaption
    \label{fig:1f}
  \end{subfigure}
  \begin{subfigure}{0pt}
    \phantomcaption
    \label{fig:1g}
  \end{subfigure}
  \begin{subfigure}{0pt}
    \phantomcaption
    \label{fig:1h}
  \end{subfigure}
\end{figure*}

\section{Results and discussion}

\subsection{Crystal structure and nanofabrication robustness} 
The Heusler compound \ch{Co2MnSi} is characterized by its half-metallic properties with a full minority spin band gap of $\Delta E\simeq 0.7\,\mathrm{eV}$ \supercite{guillemardUltralowMagneticDamping2019, demeloUnveilingTransportProperties2021} around the Fermi energy, implying a $100\%$ spin polarization \supercite{jourdanDirectObservationHalfmetallicity2014} and an associated ultralow Gilbert damping \supercite{guillemardUltralowMagneticDamping2019}. Key to achieving these desirable properties is a precise stoichiometry \supercite{guillemardPolycrystallineCo2MnbasedHeusler2019, guillemardEngineeringCo2MnAlxSi1Heusler2020} and an intact Heusler crystal structure presenting the chemical L2\textsubscript{1} order as illustrated in \Cref{fig:1a}. Such high-quality L2\textsubscript{1}-ordered \ch{Co2MnSi} thin films can be grown by molecular beam epitaxy (MBE) \supercite{guillemardIssuesGrowingHeusler2020}. We follow the procedure by Guillemard \textit{et al.} \supercite{guillemardIssuesGrowingHeusler2020} to grow an epitaxial film stack of \ch{MgO}\textbackslash\ch{Co2MnSi}\textbackslash\ch{MgO}\textbackslash\ch{Al} on a double-side polished \ch{MgO}(001) substrate as illustrated in \Cref{fig:1b}. Reflection high energy electron diffraction (RHEED) reveals the appearance of $\nicefrac{1}{2}$-streaks when probing along the $\langle110\rangle_\mathrm{\ch{Co2MnSi}}$ direction as shown in \Cref{fig:1c}, giving an \textit{in situ} indication of a high chemical ordering degree \supercite{neggacheTestingEpitaxialCo15Fe15Ge0012014}. To ensure consistent comparability throughout the study, all subsequent steps are carried out on pieces of the same sample. 

The crystal structure is investigated further (\textit{ex situ}) on a \ch{Ga+} focused ion beam (FIB) cut lamella by high resolution transmission electron microscopy (HRTEM), consolidating the epitaxial growth in the desired Heusler structure: The HRTEM analysis across the lamella confirms a homogeneous film stack, represented by the micrograph in \Cref{fig:1d}. The high angle annular dark field scanning transmission electron microscopy (HAADF-STEM) analysis reveals the characteristic intensity pattern as shown in \Cref{fig:1e} evidencing a predominant chemical L2\textsubscript{1} order \supercite{hasnipEffectCobaltSublatticeDisorder2014, guillemardIssuesGrowingHeusler2020} across the lamella.

We fabricate waveguide structures with feature sizes of $5\,\mathrm{\upmu m}$ down to $50\,\mathrm{nm}$ in a common top-down nanostructuring approach using electron beam lithography (EBL) and a resist mask \ch{Ar+} ion beam etching (IBE) procedure. Such top-down patterning bears the risk of inducing structural defects through ion collisions, as well as potential contamination by ion implantation, both known to harm the desired properties of the half-metallic Heusler compound \supercite{gaierInfluenceL21Ordering2008, abdallahEvolutionMagneticProperties2017, mantionInfluenceGaMilling2022}. We quantify the impact of nanofabrication on the Heusler crystal structure by investigating FIB cut lamellas of cross-sections of the patterned structures in HRTEM and HAADF-STEM. The post-nanofabrication analysis confirms the preserved crystal structure, with representative micrographs in \Cref{fig:1f,fig:1g} showing the most critical case of the smallest patterned structures with $50\,\mathrm{nm}$ nominal size. Remarkably, the Heusler crystal proves robust upon nanofabrication even to such small feature sizes, whereby an intact L2\textsubscript{1} ordering is evidenced even in the waveguide edges most exposed to potential ion milling damage, as shown in the zoom-in in \Cref{fig:1h}.

\subsection{Magnetic properties and magnetocrystalline anisotropy} 
\begin{figure*}[ht!]
  \centering
  \includegraphics{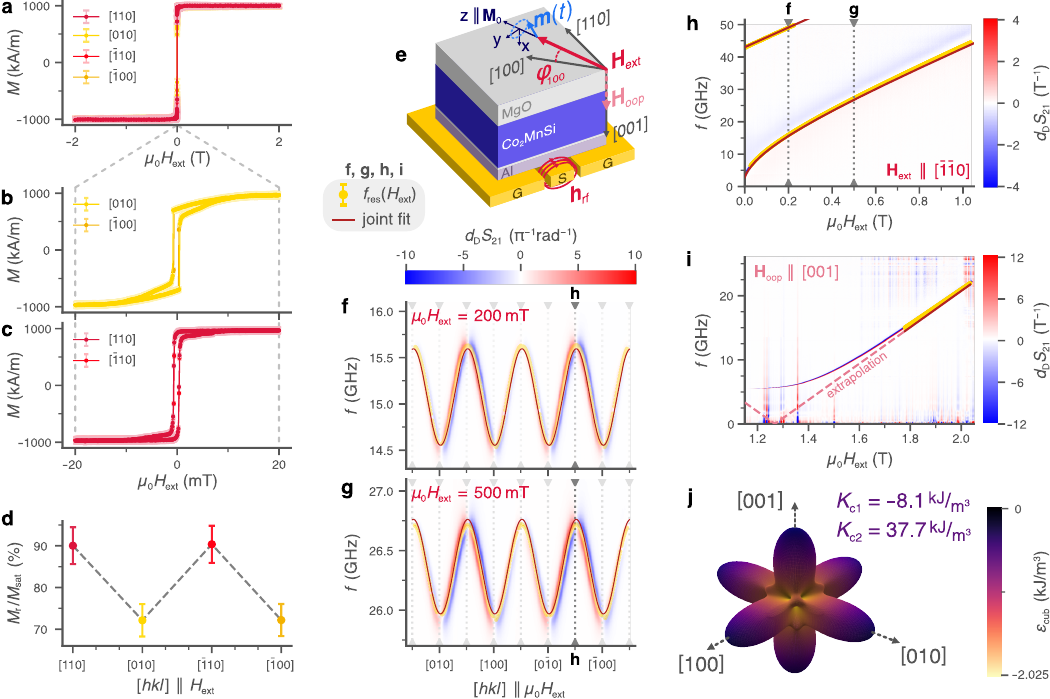}
  \caption[Cubic magnetocrystalline anisotropy in \ch{Co2MnSi}.]{
      \textbf{Cubic magnetocrystalline anisotropy in \ch{Co2MnSi}.} \textit{Note: All lattice directions $[hkl]$ in this figure are given in the \ch{Co2MnSi} crystal coordinates defined in \Cref{fig:1a}.} 
      \textbf{(a)} Hysteresis of $M(H_\mathrm{ext})$ from VSM with $\left|\mu_0\eqsHext\right| \leq 2\,\mathrm{T}$ applied in the film plane along the indicated lattice direction. 
      \textbf{(b, c)} Small-field hysteresis $M(H_\mathrm{ext})$ from VSM with $\left|\mu_0\eqsHext\right| \leq 20\,\mathrm{mT}$ applied in the film plane along the indicated lattice direction. 
      \textbf{(d)} Relative remanence magnetization $\eqsMr\,/\,\eqsMs$, with $\eqsMr$ extracted from \textbf{b, c} and $\eqsMs$ from \textbf{a}. 
      \textbf{(e)} Schematic illustration of the rotational BB-FMR geometry with the relevant crystal directions and the dynamic coordinates $x, y, z$. 
      \textbf{(f, g, h, i)} Real part of the $S_\mathrm{21}$ signal after the derivative divide background removal procedure $d_\mathrm{D}S_\mathrm{21}$ (color maps), extracted resonance frequencies $f_\mathrm{res}\left(H_\mathrm{ext}\right)$ (yellow markers) from spectral fits to real and imaginary part, and joint fit result (red line) as a function of the in-plane orientation for a fixed bias field $\mu_0\eqsHext=200\,\mathrm{mT}$ (\textbf{f}) and $\mu_0\eqsHext=500\,\mathrm{mT}$ (\textbf{g}), as well as for varying bias fields along fixed geometries (\textbf{h}, \textbf{i}). The dashed lines in \textbf{f, g} indicate all directions along which field-dependent datasets were recorded, with the example for $\mathbf{H}_\mathrm{ext}\parallel\left[\bar{1}\bar{1}0\right]$ $\left(\varphi_\mathrm{100}=5\cdot\pi/4\right)$ shown in \textbf{h}. The match to the out-of-plane FMR dataset is evaluated for $1.8\,\mathrm{T} < \mu_0 \eqsHext$, an extrapolation to the remaining data range is indicated by the dashed line in muted red. 
      \textbf{(j)} 3D surface plot of the resulting $\varepsilon_\mathrm{cub}$ in the crystal coordinates. 
  }
  \label{fig:2_anisotropy}

  \begin{subfigure}{0pt}
    \phantomcaption
    \label{fig:2a}
  \end{subfigure}
  \begin{subfigure}{0pt}
    \phantomcaption
    \label{fig:2b}
  \end{subfigure}
  \begin{subfigure}{0pt}
    \phantomcaption
    \label{fig:2c}
  \end{subfigure}
  \begin{subfigure}{0pt}
    \phantomcaption
    \label{fig:2d}
  \end{subfigure}
  \begin{subfigure}{0pt}
    \phantomcaption
    \label{fig:2e}
  \end{subfigure}
  \begin{subfigure}{0pt}
    \phantomcaption
    \label{fig:2f}
  \end{subfigure}
  \begin{subfigure}{0pt}
    \phantomcaption
    \label{fig:2g}
  \end{subfigure}
  \begin{subfigure}{0pt}
    \phantomcaption
    \label{fig:2h}
  \end{subfigure}
  \begin{subfigure}{0pt}
    \phantomcaption
    \label{fig:2i}
  \end{subfigure}
  \begin{subfigure}{0pt}
    \phantomcaption
    \label{fig:2j}
  \end{subfigure}
\end{figure*}

Maintaining an intact crystal lattice throughout nanofabrication primarily implies a preservation of magnetic properties, which are inherently tied to the crystal lattice and its symmetry. We quantify the magnetic properties of the full film prior to nanofabrication from a joint evaluation of vibrating sample magnetometry (VSM) and broadband ferromagnetic resonance spectroscopy (BB-FMR) data as presented in \Cref{fig:2_anisotropy}. The saturation magnetization is extracted from the VSM measurements of $M\left(\eqsHext\right)$ with $\left|\mu_0\eqsHext\right| \leq 2\,\mathrm{T}$ in \Cref{fig:2a}, and is consistent with literature as indicated in \Cref{tbl:parameters}. A non-negligible magnetocrystalline anisotropy effect is obtained for the small-field hystereses of $M\left(\eqsHext\right)$ with $\left|\mu_0\eqsHext\right| \leq 20\,\mathrm{mT}$ presented in \Cref{fig:2b,fig:2c}, as highlighted by the remanence magnetization $\eqsMr\,/\,\eqsMs$ in \Cref{fig:2d}. We observe the expected cubic symmetry of the magnetocrystalline anisotropy in \ch{Co2MnSi} with hard axes along the principal lattice directions $\langle100\rangle_\mathrm{\ch{Co2MnSi}}$ \supercite{gaierInfluenceL21Ordering2008, kubotaStructureExchangeStiffness2009, abdallahCorrigendumEvolutionMagnetic2017}. 

This is further quantified by in-plane rotational BB-FMR spectroscopy as illustrated in \Cref{fig:2e}. A joint evaluation of the in-plane ferromagnetic resonance $f_\mathrm{ip}\left(\eqsHext, \varphi_\mathrm{100}\right)$ is performed on two rotational FMR datasets for fixed fields $\mu_0\eqsHext=200\,\mathrm{mT}$ and $500\,\mathrm{mT}$ shown in \Cref{fig:2f,fig:2g}, as well as eight field-dependent datasets along the high-symmetry directions, i.e. for fixed orientations $\varphi_\mathrm{100}=m\cdot\pi/4$ with $m\in[0,7]_\mathbb{N}$, an example of which is shown in \Cref{fig:2h}. The anisotropic thin film FMR is modelled following \supercite{kalinikosTheoryDipoleexchangeSpin1986,kalinikosDipoleexchangeSpinWave1990} by 
\begin{equation}
    \omega_\mathrm{ip} = \sqrt{\left( \eqsomegaext + \eqsomegaexch + \eqsomegaM N_{yy}^\mathrm{ani} \right) \cdot \left(\eqsomegaext + \eqsomegaexch + \eqsomegaM + \eqsomegaM N_{xx}^\mathrm{ani} \right) }
    \label{eq:anisotropic_FMR_ip_Kittel}
\end{equation}
and
\begin{equation}
    \omega_\mathrm{oop} = \eqsomegaext + \eqsomegaexch + \eqsomegaM \cdot \left( \frac{2K_\mathrm{c1}}{\mu_0\eqsMs^2} + \frac{2K_\mathrm{oop}}{\mu_0\eqsMs^2}\right) - \eqsomegaM
    \label{eq:anisotropic_FMR_oop}
\end{equation}
where $\eqsomegaext = \gamma \mu_0 \eqsHext$ and $\eqsomegaM = \gamma \mu_0 \eqsMs$ with the gyromagnetic ratio $\gamma$, and $\eqsomegaexch = \gamma \cdot \nicefrac{2A_\mathrm{exch}}{\mu_0 \eqsMs^2} \cdot \nicefrac{n*\pi}{d}$ with mode number $n$, exchange constant $A_\mathrm{exch}$ and film thickness $d$. 

The anisotropy tensor elements $N^\mathrm{ani}_{xx}$ (out-of-plane dynamic coordinate, see \Cref{fig:2e}) and $N^\mathrm{ani}_{yy}$ (in-plane dynamic coordinate, see \Cref{fig:2e}) for a (001)-oriented, cubic lattice are given by
\begin{equation}
    N^\mathrm{ani}_{xx} = \frac{2K_\mathrm{c1}}{\mu_0\eqsMs^2} + \frac{2K_\mathrm{c2}}{\mu_0\eqsMs^2} \cdot \sin^2\left(\varphi_{100}\right)\cos^2\left(\varphi_{100}\right) - \frac{2K_\mathrm{oop}}{\mu_0\eqsMs^2} 
    \label{eq: Nxx} 
\end{equation}
and
\begin{equation}
    N^\mathrm{ani}_{yy} = \frac{2K_\mathrm{c1}}{\mu_0\eqsMs^2} \cdot \Big(1 - 6 \cdot \sin^2\left(\varphi_{100}\right)\cos^2\left(\varphi_{100}\right) \Big)  
    \label{eq: Nyy} 
\end{equation}
for the (001)-oriented thin film, where $\varphi_{100}$ is the angle between the static magnetization and $\left[100\right]_\mathrm{\ch{Co2MnSi}}$, $K_\mathrm{c1}$ and $K_\mathrm{c2}$ are the first- and second-order cubic anisotropy constants respectively, and $K_\mathrm{oop}$ accounts for a potential out-of-plane uniaxial anisotropy. We determine the best fit parameters from a joint evaluation of the entire dataset, in which $\eqsMs$ is fixed to the result obtained from the VSM evaluation, and the remaining free parameters in \Cref{eq:anisotropic_FMR_ip_Kittel} are shared. Additionally, we evaluate the fit of \Cref{eq:anisotropic_FMR_oop} to an out-of-plane dataset $f_\mathrm{oop}\left(\eqsHext\right)$ shown in \Cref{fig:2i}. The overall best fit is obtained for $K_\mathrm{oop}=0$ and reveals a cubic anisotropy with a non-negligible second-order contribution, i.e. for $K_\mathrm{c1}=\left(-8.1\pm0.9\right)\,\nicefrac{\mathrm{kJ}}{\mathrm{m}^3}$ and $K_\mathrm{c2}=\left(37.7\pm19.4\right)\,\nicefrac{\mathrm{kJ}}{\mathrm{m}^3}$, with the corresponding fit curves plotted in \Cref{fig:2f,fig:2g,fig:2h,fig:2i}. This second-order contribution partly counteracts the first-order term of opposite sign, and with $9/4 < \left|K_\mathrm{c2}/K_\mathrm{c1}\right| < 9$ \supercite{cullityIntroductionMagneticMaterials2009} yields local energy maxima (of the cubic magnetocrystalline anisotropy) along the $\langle111\rangle_\mathrm{\ch{Co2MnSi}}$ directions as illustrated in \Cref{fig:2j}. Hence, the $\left\{100\right\}_\mathrm{\ch{Co2MnSi}}$ planes host both the cubic anisotropy's global easy axes $\langle110\rangle_\mathrm{\ch{Co2MnSi}}$ 'separated' by its global hard axes $\langle100\rangle_\mathrm{\ch{Co2MnSi}}$. 

\setlength{\tabcolsep}{3pt} 
\begin{table}
    \caption[Key parameters from VSM and FMR characterization.]{
        \textbf{Key parameters from VSM and FMR.} Saturation magnetization\supercite{websterMagneticChemicalOrder1971, ritchieMagneticStructuralTransport2003, guillemardUltralowMagneticDamping2019, guillemardEngineeringCo2MnAlxSi1Heusler2020, demeloUnveilingTransportProperties2021} $\eqsMs$, gyromagnetic ratio\supercite{guillemardEngineeringCo2MnAlxSi1Heusler2020, demeloUnveilingTransportProperties2021} $\gamma$ and Landé factor $g$, magnetocrystalline anisotropy constants\supercite{gaierInfluenceL21Ordering2008, kubotaStructureExchangeStiffness2009, abdallahCorrigendumEvolutionMagnetic2017} $K_\mathrm{c1}$ and $K_\mathrm{c2}$, exchange constant\supercite{ritchieMagneticStructuralTransport2003, abdallahCorrigendumEvolutionMagnetic2017, kubotaStructureExchangeStiffness2009} $A_\mathrm{exch}$, inhomogeneous linewidth\supercite{guillemardEngineeringCo2MnAlxSi1Heusler2020, demeloUnveilingTransportProperties2021} $\mu_0\Delta H_0$ and Gilbert damping parameter\supercite{guillemardUltralowMagneticDamping2019, guillemardEngineeringCo2MnAlxSi1Heusler2020, demeloUnveilingTransportProperties2021} $\alpha$. 
    }
    \label{tbl:parameters}
    \centering
    \begin{tabular}{| l | r l | r l | }
        \hline 
        & \multicolumn{2}{c|}{this work} & \multicolumn{2}{c|}{reference} \\
        \hline
        \multirow{2}{9mm}{$\eqsMs$} & $996\pm5$     & $\mathrm{kA/m}$           & $977-1042$            & $\mathrm{kA/m}$ 
        \\
                                    & $4.87\pm0.04$ & $\mathrm{\upmu_B/f.u.}$   & $4.78-5.1$             & $\mathrm{\upmu_B/f.u.}$
        \\
        \hline
        $\gamma$                    & $28.4\pm0.3$  & $\mathrm{GHz/T}$          & $28.1-28.4$           & $\mathrm{GHz/T}$ 
        \\
        $g$                         & $2.03\pm0.02$ &                           & $2.01-2.03$           & 
        \\
        $K_\mathrm{c1}$             & $-8.1\pm0.9$  & $\mathrm{kJ/m^3}$         & $-3.93\,\,-\,\,-18$   & $\mathrm{kJ/m^3}$ 
        \\
        $K_\mathrm{c2}$             & $37.7\pm19.4$ & $\mathrm{kJ/m^3}$         & \multicolumn{2}{c|}{\textit{- commonly neglected -}} 
        \\
        $A_\mathrm{exch}$           & $25.5\pm0.2$  & $\mathrm{pJ/m}$           & $19-23.5$             & $\mathrm{pJ/m}$ 
        \\
        $\mu_0\Delta H_0$           & $0.92\pm0.01$ & $\mathrm{mT}$             & $0.43-3.12$           & $\mathrm{mT}$ 
        \\
        $\alpha$                    & $1.17\pm0.01$ & $\times\,10^{-3}$         & $0.46-2.1$            & $\times\,10^{-3}$ 
        \\
        \hline
    \end{tabular}
\end{table}

The key parameters from the joint evaluation of VSM and FMR data are listed in \Cref{tbl:parameters}. We obtain a general agreement to literature on comparable \ch{Co2MnSi} films \supercite{websterMagneticChemicalOrder1971, ritchieMagneticStructuralTransport2003, gaierInfluenceL21Ordering2008, kubotaStructureExchangeStiffness2009, abdallahCorrigendumEvolutionMagnetic2017, guillemardUltralowMagneticDamping2019, guillemardEngineeringCo2MnAlxSi1Heusler2020, demeloUnveilingTransportProperties2021}. We note that the in-plane FMR linewidth analysis in our monocrystalline L2\textsubscript{1}-ordered \ch{Co2MnSi} thin film does not confirm the acclaimed \supercite{yilginAnisotropicIntrinsicDamping2007} anisotropic Gilbert damping $\alpha$, as detailed in the \hyperlink{Supporting}{supplementary material}, and we therefore list the average damping parameter $\alpha$ and inhomogeneous linewidth $\mu_0\Delta H_0$ in \Cref{tbl:parameters}. Remarkably, our VSM and FMR results reveal a cubic magnetocrystalline anisotropy with a non-negligible second-order contribution in the L2\textsubscript{1}-ordered Heusler compound, with particular relevance for the effective magnetization at low external bias fields. Moreover, for the (001)-oriented \ch{Co2MnSi} film, the cubic magnetocrystalline anisotropy yields easy axes $\langle110\rangle_\mathrm{\ch{Co2MnSi}}$ and hard axes $\langle100\rangle_\mathrm{\ch{Co2MnSi}}$ in the film plane, and hence implies an expected stabilization of the magnetization dynamics for an alignment along an in-plane $\langle110\rangle_\mathrm{\ch{Co2MnSi}}$, particularly towards low bias fields. 

\subsection{Anisotropy-induced band gap and instability suppression}

\begin{figure}
  \centering
  \includegraphics{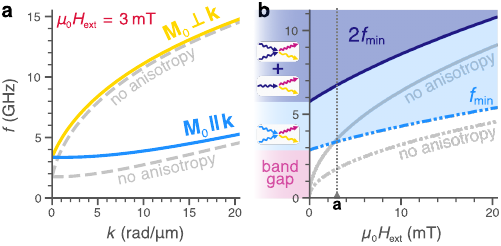}
  \caption[Anisotropy impact on the spin-wave dispersion.]{
    \textbf{Anisotropy impact on the spin-wave dispersion.} 
    \textbf{(a)} Isotropic and anisotropic spin-wave dispersion \supercite{kalinikosTheoryDipoleexchangeSpin1986,kalinikosDipoleexchangeSpinWave1990} for ${\mu_0 \eqsHext = 3\,\mathrm{mT}}$ for the $20\,\mathrm{nm}$ Co\textsubscript{2}MnSi film magnetized along an easy axis $\langle110\rangle_\mathrm{\ch{Co2MnSi}}$. 
    \textbf{(b)} Corresponding spin-wave band bottom frequency $f_\mathrm{min}\left(\eqsHext\right)$ and $2f_\mathrm{min}\left(\eqsHext\right)$ and resulting first-order and second-order nonlinear instability ranges.
  }
  \label{fig:3_calc_fmin}

  \begin{subfigure}{0pt}
    \phantomcaption
    \label{fig:3a}
  \end{subfigure}
  \begin{subfigure}{0pt}
    \phantomcaption
    \label{fig:3b}
  \end{subfigure}
\end{figure}

For an alignment along an in-plane easy axis, the cubic magnetocrystalline anisotropy shifts the spin-wave dispersion relation \supercite{kalinikosTheoryDipoleexchangeSpin1986,kalinikosDipoleexchangeSpinWave1990} to higher frequencies, as shown in \Cref{fig:3a}, and yields a non-zero band bottom frequency $f_\mathrm{min}$ even towards vanishing bias field. The resulting spin-wave band gap directly translates to an exclusion range $f_\mathrm{min} \leq f \leq 2f_\mathrm{min}$ as indicated in \Cref{fig:3b}, for an excitation in which the first-order Suhl instability, i.e. the resonant three magnon scattering $\omega_\mathrm{1}, \mathbf{k}_\mathrm{1} \rightarrow \omega_\mathrm{A}, \mathbf{k}_\mathrm{A} + \omega_\mathrm{B}, \mathbf{k}_\mathrm{B}$ is suppressed by energy conservation. Higher order processes such as the second-order Suhl instability, i.e. the resonant four magnon scattering $\omega_\mathrm{1}, \mathbf{k}_\mathrm{1a} + \omega_\mathrm{1}, \mathbf{k}_\mathrm{1b} \rightarrow \omega_\mathrm{A}, \mathbf{k}_\mathrm{A} + \omega_\mathrm{B}, \mathbf{k}_\mathrm{B}$, are not excluded but generally present higher pumping thresholds \supercite{suhlTheoryFerromagneticResonance1957}. The cubic magnetocrystalline anisotropy therefore not only results in a band gap and first-order instability suppression range, but also stabilizes the linear dynamics for higher excitation powers in this range, and hence yields a powerful toggle for linear/nonlinear device functionality control. 

\begin{figure}
  \centering
  \includegraphics{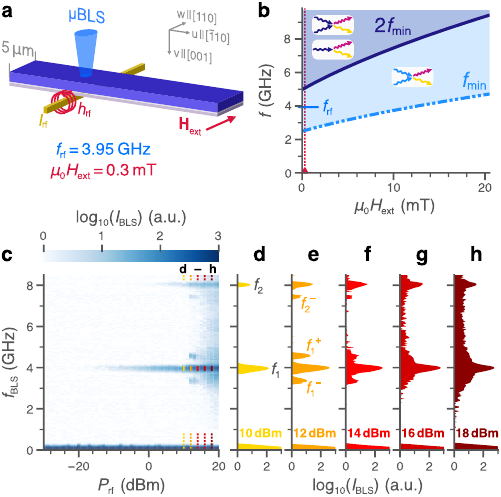}
  \caption[Anisotropy-induced instability suppression.]{
    \textbf{Anisotropy-induced instability suppression.} 
    \textbf{(a)} Schematic of the \muBLS probing. 
    \textbf{(b)} Spin-wave band bottom frequency $f_\mathrm{min}\left(\eqsHext\right)$ and $2f_\mathrm{min}\left(\eqsHext\right)$ considering the effective waveguide width $w_\mathrm{eff}=4.94\,\mathrm{\upmu m}$, and resulting first-order instability suppression range. 
    \textbf{(c)} BLS intensity spectra $I_\mathrm{BLS}\left(f_\mathrm{BLS}\right)$ as a function of the applied excitation power.  
    \textbf{(d-h)} Selected BLS spectra across the instability onset.
  }
  \label{fig:4_instability_suppression}

  \begin{subfigure}{0pt}
    \phantomcaption
    \label{fig:4a}
  \end{subfigure}
  \begin{subfigure}{0pt}
    \phantomcaption
    \label{fig:4b}
  \end{subfigure}
  \begin{subfigure}{0pt}
    \phantomcaption
    \label{fig:4c}
  \end{subfigure}
  \begin{subfigure}{0pt}
    \phantomcaption
    \label{fig:4d}
  \end{subfigure}
  \begin{subfigure}{0pt}
    \phantomcaption
    \label{fig:4e}
  \end{subfigure}
  \begin{subfigure}{0pt}
    \phantomcaption
    \label{fig:4f}
  \end{subfigure}
  \begin{subfigure}{0pt}
    \phantomcaption
    \label{fig:4g}
  \end{subfigure}
  \begin{subfigure}{0pt}
    \phantomcaption
    \label{fig:4h}
  \end{subfigure}
\end{figure}

This qualitative anisotropy implication can be probed by studying the nonlinear instability onset in the patterned \ch{Co2MnSi} structures, as presented in \Cref{fig:4_instability_suppression}. To this aim, we measure the power dependent dynamics of propagating spin waves by microfocused Brillouin light scattering spectroscopy\supercite{sebastianMicrofocusedBrillouinLight2015} ($\upmu\mathrm{BLS}$) as illustrated in \Cref{fig:4a}. We select the largest fabricated waveguide of $5\,\mathrm{\upmu m}$ width for this study, in order to avoid a misleading additional scattering partner exclusion induced by a width quantization. We calculate the first-order instability exclusion range in \Cref{fig:4b} by additionally considering the width quantization in the spin-wave dispersion for anisotropic thin films \supercite{kalinikosTheoryDipoleexchangeSpin1986,kalinikosDipoleexchangeSpinWave1990}, i.e. $\left|\mathbf{k}_\mathrm{ip}\right|^2=k^2 + k^2_\mathrm{w}$ with a fixed wavevector component along the short axis of the waveguide $k_\mathrm{w} = \nicefrac{m\cdot\pi}{w_\mathrm{eff}}$ for mode number $m$ and accounting for dipolar pinning effects through the effective width \supercite{pirroInterferenceCoherentSpin2011,bracherParallelPumpingMagnon2017,
wangSpinPinningSpinWave2019} $w_\mathrm{eff} = 4.94\,\mathrm{\upmu m}$. 

\Cref{fig:4c} presents the BLS intensity spectra $I_\mathrm{BLS}\left(f_\mathrm{BLS}\right)$ as a function of the applied excitation power $P_\mathrm{rf}$ for an excitation at $f_\mathrm{rf}=3.95\,\mathrm{GHz}$ in the near-zero bias field $\mu_0\eqsHext = 0.3\,\mathrm{mT}$. As the excitation power is increased, the direct excitation $f_\mathrm{1}=f_\mathrm{rf}$ is increasingly populated, and additionally for $P_\mathrm{rf}\geq0\,\mathrm{dBm}$ $\left(1\,\mathrm{mW}\right)$ a population of the second harmonic $f_\mathrm{2}=2\cdot f_\mathrm{rf}$ is observed linked to the elliptical precession at low fields \supercite{bracherParallelPumpingMagnon2017}. The first and second harmonic are the only detected modes for $P_\mathrm{rf}<12\,\mathrm{dBm}$ $\left(16\,\mathrm{mW}\right)$ as marked in the spectrum in \Cref{fig:4d}. At $P_\mathrm{rf}=12\,\mathrm{dBm}$, the onset of the second-order instability is observed with a characteristic population of secondary modes $f^{-}_\mathrm{1}$ and $f^{+}_\mathrm{1}$ symmetric around $f_\mathrm{1}$, as indicated in \Cref{fig:4e}. Additionally, a population of the secondary mode $f^{-}_\mathrm{2}=f_\mathrm{1}+f^{-}_\mathrm{1}$ is observed in \Cref{fig:4e}, possibly linked to a confluence of $f_\mathrm{1}$ and $f^{-}_\mathrm{1}$, or to an instability of the second harmonic $f_\mathrm{2}$ with the associated scattering partner $f^{+}_\mathrm{2}=f_\mathrm{1}+f^{+}_\mathrm{1}$ exceeding the detection range. Analogously, a characteristic population of secondary modes, $f^{-}_\mathrm{1/2}$ and $f^{+}_\mathrm{1/2}$, symmetric around $f_\mathrm{1/2}=\nicefrac{1}{2}\cdot f_\mathrm{1}\approx 2\,\mathrm{GHz}$ would indicate a first-order instability usually expected at lower pumping thresholds \supercite{suhlTheoryFerromagneticResonance1957}. The absence of signal symmetric around $f_\mathrm{1/2}$ therefore confirms the expected first-order instability suppression. Further increasing the excitation power yields an increasing energy redistribution in the system, broadening the spin-wave spectra in \Cref{fig:4f,fig:4g,fig:4h}. For $P_\mathrm{rf} > 16\,\mathrm{dBm}$, a continuous background population is detected for $f \gtrsim 2.5\,\mathrm{GHz}$ as seen in \Cref{fig:4h}, in agreement with the expected band bottom $f_\mathrm{min}$ from \Cref{fig:4b}. 

We conclude that the measurement in \Cref{fig:4_instability_suppression} thus confirms the predicted band gap up to $f_\mathrm{min}\approx 2.5\,\mathrm{GHz}$ for the near-zero bias field $\mu_0\eqsHext = 0.3\,\mathrm{mT}$, as well as the consequent suppression of the first-order instability for ${f_\mathrm{rf} < 2f_\mathrm{min}}$, showcasing the significant impact of the maintained cubic magnetocrystalline anisotropy on the nonlinear dynamics in patterned devices. 

\subsection{Anisotropy-stabilized propagation at low bias fields}

\begin{figure*}
  \centering
  \includegraphics{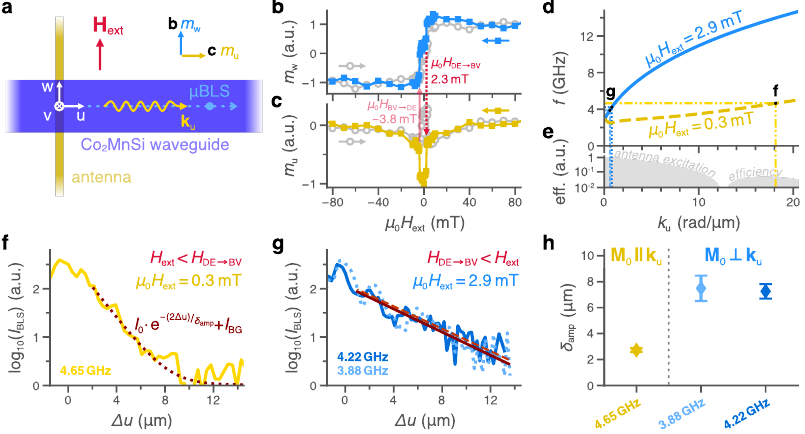}
  \caption[Anisotropy-stabilized spin-wave propagation in the low-field regime.]{
    \textbf{Anisotropy-stabilized spin-wave propagation in the low-field regime.} 
    \textbf{(a)} Top view of the measurement geometry in $\upmu\mathrm{BLS}$ and Kerr Microscopy. 
    \textbf{(b, c)} Kerr microscopy hysteresis for $\mathbf{H}_\mathrm{ext}\parallel\hat{\mathbf{w}}$ of the normalized magnetization components $m_\mathrm{w}$ and $m_\mathrm{u}$. 
    \textbf{(d, e)} DE dispersion for $\mu_0 H_\mathrm{ext}=2.9\,\mathrm{mT}>\mu_0 H_\mathrm{DE\rightarrow BV}$ and BV dispersion for $\mu_0 H_\mathrm{ext}=0.3\,\mathrm{mT}<\mu_0 H_\mathrm{DE\rightarrow BV}$ and underlying antenna excitation efficiency spectrum. 
    \textbf{(f)} Intensity decay trace for $\mu_0 H_\mathrm{ext}=0.3\,\mathrm{mT}<\mu_0 H_\mathrm{DE\rightarrow BV}$ for an excitation with $f_\mathrm{rf}=4.65\,\mathrm{GHz}$ (yellow solid line) in the linear regime ($P_\mathrm{rf}=0\,\mathrm{dBm}$), with the exponential decay fit $I=I_0 \cdot \mathrm{e}^{\nicefrac{-2\Delta u}{\delta_\mathrm{amp}}}+I_\mathrm{BG}$ (dotted red line). 
    \textbf{(g)} Intensity decay traces for $\mu_0 H_\mathrm{ext}=2.9\,\mathrm{mT}>\mu_0 H_\mathrm{DE\rightarrow BV}$ for an excitation with $f_\mathrm{rf}=4.22\,\mathrm{GHz}$ (solid line, solid fitline) and $f_\mathrm{rf}=3.88\,\mathrm{GHz}$ (dotted line, dashed fitline) in the linear regime ($P_\mathrm{rf}=-8\,\mathrm{dBm}$). 
    \textbf{(h)} Extracted amplitude decay lengths $\delta_\mathrm{amp}$ showing an increase by $\sim$$\times3$ from (f) to (g) indicating the DE propagation ($\mathbf{M}_0\perp\mathbf{k}_\mathrm{u}$) for $\mu_0 H_\mathrm{ext}=2.9\,\mathrm{mT}$.
  }
  \label{fig:5_propagation}

  \begin{subfigure}{0pt}
    \phantomcaption
    \label{fig:5a}
  \end{subfigure}
  \begin{subfigure}{0pt}
    \phantomcaption
    \label{fig:5b}
  \end{subfigure}
  \begin{subfigure}{0pt}
    \phantomcaption
    \label{fig:5c}
  \end{subfigure}
  \begin{subfigure}{0pt}
    \phantomcaption
    \label{fig:5d}
  \end{subfigure}
  \begin{subfigure}{0pt}
    \phantomcaption
    \label{fig:5e}
  \end{subfigure}
  \begin{subfigure}{0pt}
    \phantomcaption
    \label{fig:5f}
  \end{subfigure}
  \begin{subfigure}{0pt}
    \phantomcaption
    \label{fig:5g}
  \end{subfigure}
  \begin{subfigure}{0pt}
    \phantomcaption
    \label{fig:5h}
  \end{subfigure}
\end{figure*}

A typical interest in anisotropic media is the potential to stabilize an otherwise less favourable configuration. For the \ch{Co2MnSi} waveguides along $\hat{\mathbf{u}}\parallel\langle110\rangle_{\ch{Co2MnSi}}$, the long axis $\hat{\mathbf{u}}$ as well as the short axis $\hat{\mathbf{w}}$ both correspond to an easy axis of the magnetocrystalline anisotropy, separated by the hard axes along $\langle100\rangle_{\ch{Co2MnSi}}$. The latter yield an energy barrier to be overcome for switching from one configuration to another, i.e. switching between the Damon-Eshbach (DE) configuration ($\mathbf{M}_0\parallel\hat{\mathbf{w}}$) and the backward volume (BV) configuration ($\mathbf{M}_0\parallel\hat{\mathbf{u}}$) for a spin-wave propagation $\mathbf{k}\parallel\hat{\mathbf{u}}$ along the long axis. Hence, the magnetocrystalline anisotropy results in a stabilization of the magnetization in either geometry (DE as well as BV), which is particularly relevant for the dynamics at low external bias fields. We study the anisotropy-induced stabilization of the low-field regime by Kerr microscopy \supercite{soldatovSelectiveSensitivityKerr2017,mccordPerspectivesMagnetoopticalMicroscopy2025} and $\upmu\mathrm{BLS}$ as presented in \Cref{fig:5_propagation}. 

We first quantify the switching fields for the $5\,\mathrm{\upmu m}$ wide\linebreak[4] \ch{Co2MnSi} waveguide from Kerr microscopy measurements performed in the same field configuration $\mathbf{H}_\mathrm{ext}\parallel\hat{\mathbf{w}}$, as illustrated in\linebreak[4] \Cref{fig:5a}. The extracted hysteresis is shown in \Cref{fig:5b,fig:5c}. Upon saturation, the applied field can be reduced down to a minimal field of ${|\mu_0 \eqsHext| \geq |\mu_0 H_\mathrm{DE\rightarrow BV}| = 2.3\pm0.3\,\mathrm{mT}}$ sufficient to keep the waveguide in the DE configuration, as marked by the dashed arrow in \Cref{fig:5b,fig:5c}. A minimal field of $|\mu_0 \eqsHext| \geq |\mu_0 H_\mathrm{BV\rightarrow DE}| = 3.8\pm0.3\,\mathrm{mT}$ is required to re-magnetize the waveguide in the DE configuration as marked by the solid arrow accordingly. For comparison, the theoretical width-induced demagnetization field ${\mu_0 \eqsHwidth=-\mu_0 \eqsMs N_{ww}}$ with $N_{ww} \left(\Delta w\right) = \nicefrac{1}{\pi}\cdot\left(\mathrm{arctan}\left(\nicefrac{d}{2\cdot\Delta w + w}\right) - \mathrm{arctan}\left(\nicefrac{d}{2\cdot\Delta w - w}\right)\right)$ yields $\mu_0 \eqsHwidth=-3.19\,\mathrm{mT}$ at the center ${\Delta w=0}$ of the waveguide with ${w=5\,\mathrm{\upmu m}}$ and ${d=20\,\mathrm{nm}}$. The Kerr hysteresis loops thus confirm the anisotropy-induced stabilization of the DE and BV configuration with ${\left|H_\mathrm{DE\rightarrow BV}\right| < \left|\eqsHwidth\right| < \left|H_\mathrm{BV\rightarrow DE}\right|}$. 

We then quantify the resulting propagation dynamics in the low field regime by investigating the spin-wave decay. We probe the BV configuration for $\mu_0 \eqsHext = 0.3\,\mathrm{mT}$, i.e. the near-remanence case as chosen before (\Cref{fig:4c,fig:4d,fig:4e,fig:4f,fig:4g,fig:4h}), with the corresponding dispersion relation (dashed line) shown in \Cref{fig:5d}. Subsequently, we saturate the waveguide in the DE configuration by applying $\mu_0 \eqsHext = 200\,\mathrm{mT}\gg \mu_0 H_\mathrm{BV\rightarrow DE}$ and then reduce the field down to $\mu_0 \eqsHext = 2.9\,\mathrm{mT}$, i.e. $\left|H_\mathrm{DE\rightarrow BV}\right| < \left|\eqsHext\right| < \left|\eqsHwidth\right| < \left|H_\mathrm{BV\rightarrow DE}\right|$ with the corresponding dispersion relation (solid line) in \Cref{fig:5c}. In both cases, we measure the spin-wave intensity by $\upmu\mathrm{BLS}$ probing along the center line of the waveguide ($\Delta w=0$) as a function of the distance $\Delta u$ as illustrated in \Cref{fig:5a}. The most efficiently excited propagating modes result from a combination of the dispersion relation (largest group velocity $\nu_\mathrm{g}=\nicefrac{\partial\omega}{\partial k_\mathrm{u}}$) in \Cref{fig:5d} and the calculated antenna excitation efficiency spectrum in \Cref{fig:5e}, and are identified from a preliminary radiofrequency sweep evaluating the maximal intensity at $\Delta u = 2\,\mathrm{\upmu m}$. The measured decay traces for the expected BV geometry $\mathbf{M}_0\parallel\mathbf{k}_\mathrm{u}$ with $\mu_0 \eqsHext = 0.3\,\mathrm{mT}$ and the expected DE geometry $\mathbf{M}_0\perp\mathbf{k}_\mathrm{u}$ with $\mu_0 \eqsHext = 2.9\,\mathrm{mT}$ are shown in \Cref{fig:5f,fig:5g} respectively. We extract the exponential amplitude decay length $\delta_\mathrm{amp}$ shown in \Cref{fig:5h} from the intensity decay fits following \supercite{heinzPropagationSpinWavePackets2020} 
\begin{equation}
    I_\mathrm{BLS}\left(\Delta u\right)=I_0\cdot\mathrm{exp}\left(-\frac{2\cdot \Delta u}{\delta_\mathrm{amp}}\right) + I_\mathrm{BG}
\end{equation}
where $I_0$ accounts for the initial intensity and $I_\mathrm{BG}$ for the background intensity due to thermal spin waves or noise. 

For the BV decay in \Cref{fig:5f}, we obtain $\delta_\mathrm{amp}=2.7\pm0.2\,\mathrm{\upmu m}$ in the linear regime with $P_\mathrm{rf}=0\,\mathrm{dBm}$. For comparison, the expected theoretical value can be calculated by \supercite{heinzPropagationSpinWavePackets2020}
\begin{equation}
    \delta_\mathrm{theo} = \tau \cdot \nu_\mathrm{g}
    \label{eq:decay_length}
\end{equation}
with the spin-wave group velocity $\nu_\mathrm{g}=\nicefrac{\partial\omega}{\partial k_\mathrm{u}}$ and 
\begin{equation}
    \tau^{-1} = \left(\alpha\cdot\omega + \frac{\gamma\cdot \mu_0\Delta H_0}{2}\right) \cdot \frac{\partial\omega}{\partial\omega_\mathrm{H}}
    \label{eq:lifetime}
\end{equation}
defining the spin-wave lifetime $\tau$. Assuming unaffected key parameters $\alpha$, $\mu_0\Delta H_0$ and $\gamma$ as identified by the FMR characterization and listed in \Cref{tbl:parameters}, we obtain $\delta_\mathrm{theo}=2.76\,\mathrm{\upmu m}$ for the BV case in \Cref{fig:5f}, which is in agreement with the experimentally extracted value and another indication for the unaffected material parameters upon nanofabrication. 

For the decay in the increased field $\mu_0 \eqsHext = 2.9\,\mathrm{mT}$ in \Cref{fig:5g}, we obtain the significantly increased decay lengths $\delta_\mathrm{amp,\,3.88\,GHz}=7.5\pm0.9\,\mathrm{\upmu m}$ and $\delta_\mathrm{amp,\,4.22\,GHz}=7.3\pm0.6\,\mathrm{\upmu m}$, amounting to nearly triple the BV decay length as shown in \Cref{fig:5h}. This is a clear indication for the expected DE configuration with an improved propagation linked to the increased group velocity. 
Remarkably, the measurements confirm the DE configuration for an operation in an external field $\eqsHext < -\eqsHwidth$, i.e. for which the hypothetical case of a negligible magnetocrystalline anisotropy would otherwise yield a BV configuration with significantly reduced spin-wave propagation. 

We note that the extracted decay lengths from \Cref{fig:5g} amount to only $36\%$ of the respective expected values $\delta_\mathrm{theo,\,3.88\,GHz}=20.7\,\mathrm{\upmu m}$ and $\delta_\mathrm{theo,\,4.22\,GHz}=20.1\,\mathrm{\upmu m}$, which we attribute to inhomogeneities due to dipolar pinning along the edges leading to an increased dissipation. However, we emphasize that focus must not be on this discrepancy, but instead on the confirmation of the otherwise impossible DE configuration at this low bias field. Moreover, we note that these low-bias-field decay lengths $\delta_\mathrm{amp,\,3.88\,GHz}$ and $\delta_\mathrm{amp,\,4.22\,GHz}$ extracted from \Cref{fig:5g} are similar to the range of decay lengths $\delta_\mathrm{amp}=8.7-11.9\,\mathrm{\upmu m}$ reported for a comparable Co\textsubscript{2}Mn\textsubscript{0.6}Fe\textsubscript{0.4}Si waveguide in a significantly larger bias field $\mu_0 \eqsHext = 40\,\mathrm{mT}$ \supercite{sebastianLowdampingSpinwavePropagation2012}, and well above commonly reported high-field values $\delta_\mathrm{amp}<6\,\mathrm{\upmu m}$ for similarly dimensioned \ch{Ni81Fe19} waveguides \supercite{pirroInterferenceCoherentSpin2011, demidovModeInterferencePeriodic2008}. 

The decay measurements in \Cref{fig:5_propagation} therefore not only showcase the maintained key parameters and exploitable anisotropy-induced stabilization of the BV and DE dynamics, but also highlight the superior characteristics of the half-metallic \ch{Co2MnSi} compared to common metallic ferromagnets allowing for a decent propagation performance even at such low bias fields. 

\section{Conclusions}
To conclude, we establish the Heusler compound \ch{Co2MnSi} as a robust and scalable half-metallic platform for nonlinear magnonic applications, particularly towards vanishing bias fields. We demonstrate that top-down nanofabrication preserves the L2\textsubscript{1}-ordered crystal structure and the associated magnetic key parameters, addressing a central challenge for half-metallic \ch{Co2MnSi}-based nanodevices.  

We further show that the intrinsic cubic magnetocrystalline anisotropy presents a significant second-order contribution and plays a decisive role in the resulting spin-wave dynamics. In patterned waveguides, this anisotropy gives rise to a GHz-scale band gap and suppresses first-order nonlinear instabilities, while simultaneously enabling stabilized operation in the Damon–Eshbach configuration at otherwise insufficient bias fields. 

Our results highlight the cubic anisotropy as a powerful resource for controlling linear and nonlinear dynamics in half-metallic systems. More broadly, we demonstrate that L2\textsubscript{1}-order-preserving nanofabrication enables the retention of the characteristic material properties of \ch{Co2MnSi} at the nanoscale, and explicitly confirm the direct link between preserved crystal structure and magnetic functionality. This establishes a key milestone towards the realization of fully spin-polarized, low-dissipation nanoscale device architectures. 

\section{Methods}
\subsection{A. Epitaxial film growth}
The thin film sample was grown by molecular beam epitaxy (MBE) following the procedure established by Guillemard \textit{et al.} \supercite{guillemardIssuesGrowingHeusler2020} on a COMPACT 21 EB 200 MBE system from RIBER operating with a base pressure of $4\times10^{-9}\,\mathrm{Pa}$ ($4\times10^{-11}\,\mathrm{mbar}$). The system is equipped with 3 multi-pocket electron guns (in this study used for the evaporation of \ch{MgO}, \ch{Co} and \ch{Si}) and 6 effusion cells (used for \ch{Mn} and \ch{Al}). A set of 4 quartz microbalances with an INFICON IC6 controller allow for a precise calibration and real time stoichiometry control by monitoring the individual deposition flux rates. All deposition steps of the epitaxial growth process were monitored \textit{in situ} by Reflection High Energy Electron Diffraction (RHEED). Prior to the deposition process, a $500\,\mathrm{\upmu m}$ thick and $30 \times 30\,\mathrm{mm}^2$ large \ch{MgO}(001) substrate was outgassed \textit{in situ} at $T_\mathrm{pyro}\approx970\,\mathrm{K}$ (measured with a pyrometer focused on the sample surface using an arbitrary chosen emissivity equal to 0.85). Right afterwards, a $10\,\mathrm{nm}$ thick \ch{MgO} buffer layer was deposited at $T_\mathrm{pyro}\approx900\,\mathrm{K}$ in order to smoothen the surface and cover potential contaminations of the substrate. The sample was left to cool down, and subsequently the $20\,\mathrm{nm}$ thick \ch{Co2MnSi} film was deposited at $T_\mathrm{pyro}\approx670\,\mathrm{K}$ under rotation ensuring a homogeneous deposition monitored not only by RHEED but also the deposition rates detected by the quartz microbalances. 
Finally, an \ch{MgO} layer of $1\,\mathrm{nm}$ nominal thickness was deposited to ensure a symmetric top- and bottom interface of the Heusler layer, and an \ch{Al} layer of $5\,\mathrm{nm}$ thickness as a final capping. 

\subsection{B. Top-down nanostructuring}
The $30\times30\,\mathrm{mm}^2$ sample was cut to 9 pieces of $10 \times 10\,\mathrm{mm}^2$ to allow for thin film reference measurements, technical structuring tests, the nanofabrication impact study by TEM as well as a final device fabrication. For the latter two, a full nanostructuring process was carried out following a typical resist mask ion beam milling procedure. In a first step, the sample pieces were sonicated sequentially in acetone and isopropanol, and subsequently blow-dried with \ch{N2}. The negative tone resist ma-N 2403 was used in a typical EBL step to fabricate the milling mask for the subsequent \ch{Ar+} IBE process. Successive \ch{Ar+} ion milling was carried out at an angle of incident of $70^\circ$, $20^\circ$ and $70^\circ$ to the film normal. For the nanofabrication impact study by TEM, arrays of waveguides of various widths between $w = 50\,\mathrm{nm}$ and $w = 300\,\mathrm{nm}$ oriented along the $\langle110\rangle_\mathrm{\ch{Co2MnSi}}$ directions were fabricated and subsequently a cross-section was cut by FIB for the investigation in TEM. Hence, for the final spin-wave studies additional waveguides were patterned on a different piece, equally oriented along the $\langle110\rangle_\mathrm{\ch{Co2MnSi}}$ directions and following the same nanofabrication procedure, where additionally a $500\,\mathrm{nm}$ wide stripline antenna made from \ch{Ti}($10\,\mathrm{nm}$)/\ch{Au}($80\,\mathrm{nm}$) was structured on top of the waveguides in a subsequent EBL lift-off process using electron beam vapor deposition. 

\subsection{C. Structural characterization by transmission electron microscopy (TEM)}
For the TEM studies, cross-sections of the samples (thin film as well as patterned structures) were cut by Ga\textsuperscript{+} FIB etching. Prior to the FIB process, C was deposited on the sample to avoid a sample charging and deflection of the ion beam during the FIB process, and Pt was deposited at the lamella extraction site to protect the sample surface from the ion bombardment. The FIB lamellas were cut along the $\langle110\rangle_\mathrm{\ch{Co2MnSi}}$ directions in order to allow for an investigation of the chemical ordering in the Heusler compound \supercite{guillemardIssuesGrowingHeusler2020}. The FIB lamellas are thinned to be of approx. $50\,\mathrm{nm}$ width at the TEM probing site. TEM experiments were carried out on a JEM-ARM 200F Cold FEG TEM/STEM system operating at $200\,\mathrm{kV}$ acceleration voltage. The imaging was performed in two operation modes: (high resolution) transmission electron microscopy (TEM/HRTEM) where the sample is static and illuminated with a parallel beam, as well as scanning transmission electron microscopy (STEM) where the electron beam was focused on the sample. In the STEM mode, the dark field images were recorded using a high angle annular dark field (HAADF) detector, which allows for an imaging with a site-specific chemical contrast \supercite{treacyContrastPlatinumPalladium1978} and hence allows to determine the chemical ordering by a comparison of the relative intensities of neighbouring atomic columns \supercite{guillemardIssuesGrowingHeusler2020}. 

\subsection{D. Vibrating sample magnetometry (VSM)}
The VSM measurements were acquired on a commercial physical properties measurement system (PPMS) DynaCool base setup by Quantum Design, equipped with the VSM add-on option. For these VSM measurements, a $4\times4\,\mathrm{mm}^2$ was cut from one of the $10\times10\,\mathrm{mm}^2$ pieces. The $M(H)$ measurements were performed at a temperature of $T=300\,\mathrm{K}$, the VSM was operated with a peak amplitude of $A_\mathrm{VSM}=2\,\mathrm{mm}$ and a frequency of $f_\mathrm{VSM}=40\,\mathrm{Hz}$, resulting in a maximum acceleration of $a_\mathrm{VSM}=126.33\,\mathrm{m/s^2}$ and a maximum measurable moment of $\mathcal{M}_\mathrm{max}=44.59\,\mathrm{mJ/T}=44.59\times10^{-3}\,\mathrm{Am^2}$. The field was swept with a rate set to $r = 1\,\mathrm{mT/s}$, and data was acquired for an averaging time of $t_\mathrm{avg}=1\,\mathrm{s}$ per field point during which the field was controlled in the driven mode. Every measurement was conducted as a hysteresis loop, i.e. full $M(H)$ cycle, in order to enable a proper background removal in the data post-processing. An automated touchdown centring was performed at intervals of $\Delta t=10\,\mathrm{min}$ in order to avoid a vertical positioning error during the measurement \supercite{quantumdesignVibratingSampleMagnetometer2011}. The comparability between different measurements was ensured by a setup demagnetizing procedure to avoid field shifts arising from pinned vortices in the superconducting coils. The recorded absolute moment was post-processed considering the sample geometry to extract the magnetization plotted in \Cref{fig:2_anisotropy}. See the \hyperlink{Supporting}{supplementary material} for more details on the demagnetization procedure and the data post-processing.

\subsection{D. Broadband VNA-FMR spectroscopy (BB-FMR)}
BB-FMR was performed on one of the $10\times10\,\mathrm{mm}^2$ pieces using a rotational field setup equipped with a Keysight N5225B PNA microwave network analyser and a Caylar EA186 electromagnet. The sample was placed upside down on an impedance matched ($50\,\mathrm{\Omega}$) coplanar waveguide (CPW) and frequency spectra of the $S_\mathrm{21}$ parameter were measured with the (uncalibrated) vector network analyzer (VNA). For the in-plane geometry, the eight field-dependent data sets were recorded with a field range of $0\,\mathrm{T} \leq \mu_0 H_\mathrm{ext} \leq 1.04\,\mathrm{T}$ in steps of $\Delta \mu_0 H_\mathrm{ext} = 1\,\mathrm{mT}$, and the two rotational data sets as a full rotational scan with $0^\circ \leq \varphi_\mathrm{100} \leq 360^\circ$ in steps of $\Delta \varphi_\mathrm{100}=1^\circ$. The out-of-plane measurement was recorded in a field range of $0\,\mathrm{T} \leq \mu_0 H_\mathrm{ext} \leq 2.1\,\mathrm{T}$ in steps of $\Delta \mu_0 H_\mathrm{ext} = 1\,\mathrm{mT}$, but only the high-field data with $1.8\,\mathrm{T} < \mu_0 H_\mathrm{ext}$ was used for the evaluation. The raw VNA spectra were post-processed following the \textit{derivative divide} procedure \supercite{maier-flaigNoteDerivativeDivide2018} and evaluating the resonance curves in the resulting real part $\Re\left(d_\mathrm{D}S_\mathrm{21}\right)$ and imaginary part $\Im\left(d_\mathrm{D}S_\mathrm{21}\right)$ of the \textit{derivative divide} spectra $d_\mathrm{D}S_\mathrm{21}$. The anisotropic FMR is modelled following the model derived by Kalinikos and Slavin \supercite{kalinikosTheoryDipoleexchangeSpin1986,kalinikosDipoleexchangeSpinWave1990}. In addition to the commonly considered first-order cubic anisotropy term with $K_\mathrm{c1}$, we include $K_\mathrm{c2}$ and $K_\mathrm{oop}$ in the model, since an evaluation considering only the first-order cubic anisotropy (i.e. fixing $K_\mathrm{c2}=0$ and $K_\mathrm{oop}=0$ in \Cref{eq: Nxx}) fails to reproduce the data particularly along the $\langle110\rangle_\mathrm{\ch{Co2MnSi}}$ directions. 

We note that the similar symmetry of the contributions with $K_\mathrm{c1}$, $K_\mathrm{c2}$ and $K_\mathrm{oop}$ imposes to include at least 3 distinct directions for the anisotropy evaluation of (001)-oriented thin films. We therefore model the entire in-plane FMR dataset $f_\mathrm{ip}\left(\eqsHext,\varphi_\mathrm{100}\right)$ with shared parameters. To avoid an over-parametrization, we perform the full analysis for three distinct cases in \Cref{eq: Nxx}: (i) the common assumption of a cubic magnetocrystalline anisotropy with negligible higher order contributions, i.e. fixing $K_\mathrm{c2}=0$ and $K_\mathrm{oop}=0$, (ii) a potential non-negligible second-order contribution, fixing only $K_\mathrm{oop}=0$, and (iii) testing against a potential out-of-plane uniaxial anisotropy, by fixing $K_\mathrm{c2}=0$. Additionally, we evaluate the fit of \Cref{eq:anisotropic_FMR_oop} to an out-of-plane dataset $f_\mathrm{oop}\left(\eqsHext\right)$ to determine the overall best fit. Finally, we extract the Gilbert damping parameter $\alpha$ as well as the inhomogeneous linewidth $\mu_0 H_0$ from the resonance linewidth individually for the eight in-plane FMR datasets along fixed orientations. No significant anisotropy in the resonance linewidth is observed, and we therefore average over the eight directions to extract the overall damping contribution reported in \Cref{tbl:parameters}. Further details on the FMR evaluation are given in the \hyperlink{Supporting}{supplementary material}. 

\subsection{E. Theoretical calculations}
All theoretical calculations in this article are obtained using the spin-wave dispersion model for anisotropic thin films derived by Kalinikos and Slavin \supercite{kalinikosTheoryDipoleexchangeSpin1986,kalinikosDipoleexchangeSpinWave1990}. To incorporate the second-order contribution of the cubic anisotropic, we derive the linearized demagnetization tensor following a common approach \supercite{stancilSpinWavesTheory2009} as detailed in the \hyperlink{Supporting}{supplementary material} yielding the anisotropy tensor elements in \Cref{eq: Nxx,eq: Nyy}. For the calculations concerning the waveguide structures, we additionally incorporate the width quantization $\left|\mathbf{k}_\mathrm{ip}\right|^2=k^2 + k^2_\mathrm{w}$ with a fixed wavevector component along the short axis of the waveguide $k_\mathrm{w} = \nicefrac{m\cdot\pi}{w_\mathrm{eff}}$ for mode number $m$. Hereby, we use the effective width 
\begin{equation}
  w_\mathrm{eff}=w_\mathrm{phys}\cdot \frac{D_\mathrm{pin}}{D_\mathrm{pin} - 2} \,\, \mathrm{with} \,\, D_\mathrm{pin} = 2\pi \cdot \frac{\frac{w_\mathrm{phys}}{d}}{1+2\cdot\mathrm{ln}\left(\frac{w_\mathrm{phys}}{d}\right)}
\end{equation}
to consider dipolar pinning effects for a waveguide with physical width $w_\mathrm{phys}$ and thickness $d$. We account for the width-induced demagnetization field $\mu_0 \eqsHwidth=-\mu_0 \eqsMs N_{ww}$ with $N_{ww} \left(\Delta w\right) = \nicefrac{1}{\pi}\cdot\left(\mathrm{arctan}\left(\nicefrac{d}{2\cdot\Delta w + w}\right) - \mathrm{arctan}\left(\nicefrac{d}{2\cdot\Delta w - w}\right)\right)$. 

\subsection{F. Kerr microscopy}
Kerr microscopy was performed on a wide-field Kerr microscope by evico magnetics \supercite{soldatovSelectiveSensitivityKerr2017, mccordPerspectivesMagnetoopticalMicroscopy2025} using a microscope objective with $50\times$ magnification. The sample was placed in an external magnetic field applied along the short axis of the waveguide. Kerr hysteresis measurements were performed in two standard configurations \supercite{mccordPerspectivesMagnetoopticalMicroscopy2025}, i.e. probing the Kerr signal in the \textit{longitudinal + polar}, as well as \textit{transverse + polar} geometry. A full hysteresis for $-100\,\mathrm{mT}\leq\mu_0 \eqsHext\leq100\,\mathrm{mT}$ was recorded, with steps of $10\,\mathrm{mT}$ in the outer range, refined to steps of $1\,\mathrm{mT}$ for $-10\,\mathrm{mT}\leq\mu_0 \eqsHext\leq10\,\mathrm{mT}$ and to $0.5\,\mathrm{mT}$ for $-5\,\mathrm{mT}\leq\mu_0 \eqsHext\leq5\,\mathrm{mT}$. For each field value, 25 images were averaged for each field point and subsequently post-processed using a standard background subtraction, drift correction and compensation of parasitic Faraday rotation in the manufacturer-provided software \textit{Looper Offline} (evico magnetics). 

\subsection{G. Microfocused BLS spectroscopy}

The microfocused Brillouin light scattering experiments were performed following the principle decribed by Sebastian \textit{et al.} \supercite{sebastianMicrofocusedBrillouinLight2015}. The sample was positioned on a piezoelectric driven nanopositioning stage placed in an external magnetic field applied along the short axis of the waveguide. Prior to the low bias field measurements, a saturation cycle with a maximum field of $\mu_0 \eqsHext = 200\,\mathrm{mT}$ was performed to reduce domain pinning at the waveguide edges. A microwave probe was used to connect a microwave circuit to the patterned stripline antenna, allowing to inductively excite spin waves. The spatially-resolved BLS measurements were performed using a continuous-wave single-frequency laser operating at $457\,\mathrm{nm}$. To allow for a probing at the antenna position, the laser beam is focused from the backside (substrate side) of the sample onto the target \ch{Co2MnSi} waveguide using a compensating microscope objective with $100\times$ magnification and numerical aperture $0.85$, translating to a maximal detectable wavevector $k_\mathrm{ip}=24\,\nicefrac{\mathrm{rad}}{\mathrm{\upmu m}}$. The intensity of the inelastically back-scattered light $I_\mathrm{BLS}$ is recorded as a function of the frequency shift $f_\mathrm{BLS}$ with respect to a reference beam using a multipass tandem Fabry-Perot interferometer, and is directly proportional to the intensity of the probed spin waves.

\section*{Abbreviations}

List of abbreviations appearing in this manuscript, sorted by alphabetical order.

\noindent\textbf{BB-FMR} Broadband ferromagnetic resonance (spectroscopy)

\noindent\textbf{BLS} Brillouin light scattering (spectroscopy)

\noindent\textbf{CMFS} Co\textsubscript{2}Mn\textsubscript{0.6}Fe\textsubscript{0.4}Si

\noindent\textbf{CPW} Coplanar waveguide

\noindent\textbf{EBL} Electron beam lithography

\noindent\textbf{FIB} Focused ion beam

\noindent\textbf{FMR} Ferromagnetic resonance (spectroscopy)

\noindent\textbf{HAADF} High angle annular dark field

\noindent\textbf{HAADF-STEM} High angle annular dark field scanning transmission electron microscopy

\noindent\textbf{HRTEM} High resolution transmission electron microscopy

\noindent\textbf{IBE} Ion beam etching

\noindent\textbf{ip} in-plane

\noindent\textbf{MBE} Molecular beam epitaxy

\noindent\textbf{$\upmu$BLS} Microfocused Brillouin light scattering (spectroscopy)

\noindent\textbf{oop} out-of-plane

\noindent\textbf{RHEED} Reflection high energy electron diffraction

\noindent\textbf{STEM} Scanning transmission electron microscopy

\noindent\textbf{TEM} Transmission electron microscopy

\noindent\textbf{VNA} Vector network analyzer

\noindent\textbf{VSM} Vibrating sample magnetometry / magnetometer

\section*{Data availability}
The datasets generated and analysed during this study are available in the ZENODO repository: \url{https://doi.org/10.5281/zenodo.21297279}.

\section*{Code availability}
Code is available from the authors upon reasonable request.

\printbibliography

\section*{Acknowledgements}
The authors thank the European Research Council (ERC) for funding this work through the ERC Starting Grant 101042439 (CoSpiN), as well as the Agence Nationale de la Recherche (France) for funding this work via the contracts ANR-20-CE24-0012 (MARIN) and ANR-20-CE24-0023 (CONTRABASS). 
A.M. Friedel acknowledges support from the Franco-German University (FGU). 
The authors acknowledge the use of the facilities of the Centre de compétences de dépôts et analyses sous ultravide de nanomatériaux (CC D.a.u.m) at IJL as well as the Nanostructing Center (NSC) of the RPTU for the sample fabrication. 
The authors acknowledge the use of the facilities of the Centre de Compétences Microscopies, Microscondes et Métallographie (CC 3M) at IJL for the FIB and TEM study. 
The funders played no role in study design, data collection, analysis and interpretation of data, or the writing of this manuscript. 

\section*{Author contributions}
A.M.F., P.P., S.P.-W. and S.A. conceived the study. A.M.F. and S.A. fabricated the thin film sample and carried out the \textit{in-situ} RHEED experiments. B.H., M.B. and A.M.F. carried out the nanofabrication. A.M.F. carried out all VSM, FMR, Kerr microscopy and $\upmu$BLS experiments. J.G., A.M.F. and S.A. carried out the TEM experiments, for which S.M. prepared the FIB lamellas. A.M.F. performed the theoretical calculations. A.M.F. analyzed the data and visualized the results. A.M.F., P.P., S.P.-W. and S.A. discussed the results. P.P., S.P.-W. and S.A. acquired funding. A.M.F. wrote the original manuscript. All authors reviewed and commented on the manuscript. 

\section*{Competing interests}
All authors declare no financial or non-financial competing interests. 

\section*{Additional information}
\hypertarget{Supporting}{Supplementary material}: Expression of the linearized cubic anisotropy tensors and incorporation into the spin-wave dispersion model. Additional FMR fit cases and details on the FMR fit evaluation. FMR linewidth analysis. 

\end{document}